\documentclass[11pt,showpacs,preprintnumbers,amsmath,amssymb,prd,nofootinbib,superscriptaddress]{revtex4-2}

\usepackage{dcolumn}
\usepackage{bm}
\usepackage{ifpdf}
\usepackage{hyperref}
\usepackage{bm}
\usepackage{xcolor,color,graphicx,graphics,physics}
\usepackage[spanish,english]{babel}
\usepackage[latin1]{inputenc}
\usepackage[OT1]{fontenc}
\usepackage{latexsym,amssymb,amsmath,amsfonts, slashed}
\usepackage{makeidx}
\usepackage{epsfig,subfigure}
\usepackage{natbib}
\usepackage{epstopdf}
\usepackage{mathrsfs}
\usepackage{hyperref}
\hypersetup{colorlinks=true, linkcolor=blue, citecolor=blue, urlcolor=blue}
\usepackage{enumerate}

\usepackage{fixmath}

\everymath{\displaystyle}

\newcommand{\bea}{\begin{eqnarray}}
\newcommand{\eea}{\end{eqnarray}}

\newcommand{\orcid}[1]{\href{https://orcid.org/#1}{\includegraphics[width=10pt]{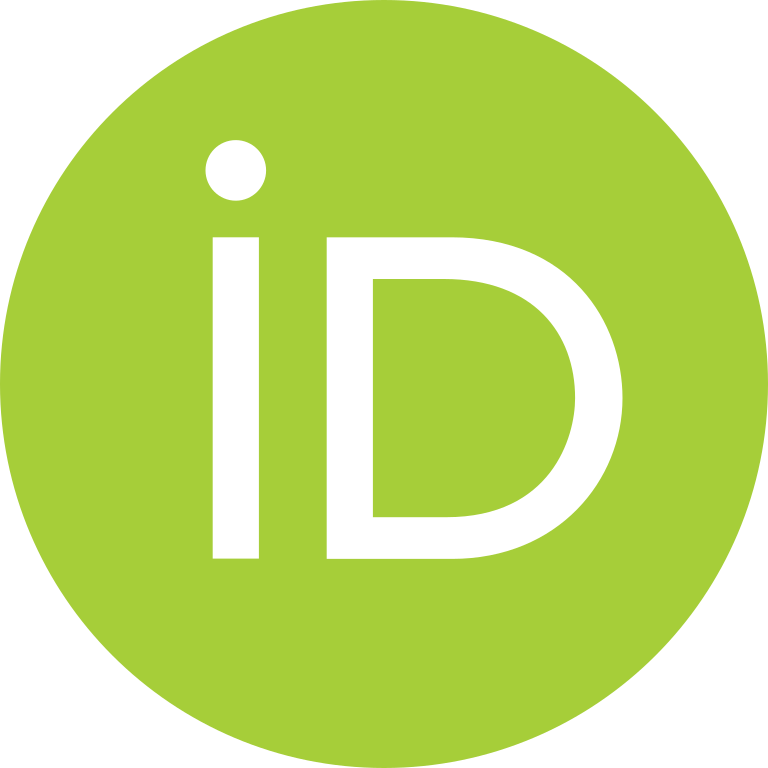}}}

\begin{document}


\title{G\"{o}del-type universes in unimodular gravity }

\author{R. Raimarda \orcid{0000-0002-4221-300X}}
\email{rigel_raimarda@fisica.ufmt.br }
\affiliation{Programa de P\'{o}s-Gradua\c{c}\~{a}o em F\'{\i}sica, Instituto de F\'{\i}sica,\\ 
Universidade Federal de Mato Grosso, Cuiab\'{a}, Brasil}

\author{A. F. Santos \orcid{0000-0002-2505-5273}}
\email{alesandroferreira@fisica.ufmt.br}
\affiliation{Programa de P\'{o}s-Gradua\c{c}\~{a}o em F\'{\i}sica, Instituto de F\'{\i}sica,\\ 
Universidade Federal de Mato Grosso, Cuiab\'{a}, Brasil}

\author{R. Bufalo \orcid{0000-0003-1879-1560}}
\email{rodrigo.bufalo@ufla.br}
\affiliation{Departamento de F\'isica, Universidade Federal de Lavras,\\
Caixa Postal 3037, 37203-202 Lavras, MG, Brazil}

\begin{abstract}

In this paper, the G\"{o}del-type universes are examined within the framework of unimodular gravity.
Since the existence of G\"{o}del solutions is intrinsically related to the presence of a cosmological constant in general relativity, one can naturally wonder how the acausal structure of the G\"{o}del solutions behaves in different cosmological scenarios.
One of the simplest, but important, frameworks within this context is the unimodular gravity, in which the cosmological constant emerges as an integration constant rather than a coupling constant.
Hence, the validity of G\"{o}del-type solutions is scrutinized within the unimodular approach, examining whether this theory can address the known issue of causality violation in G\"{o}del universes.
In detail, it is demonstrated that for certain gravitational sources, both causal and non-causal regions are permissible.

\end{abstract}

\maketitle

\section{Introduction}

There are compelling observational data indicating that the cosmological constant ($\Lambda$) is non-zero in our universe, establishing the standard cosmological model $\Lambda$CDM.
 This cosmological model describes the evolution of the Universe from its early stages to the present. It is consistent with high-precision data and offers a reasonably comprehensive account of various phenomena, including the existence and structure of the cosmic microwave background, the large-scale distribution of galaxies, the accelerating expansion of the universe, and more.
This model associates the presence of $\Lambda$ with dark energy, explaining the accelerated expansion of the universe. It also postulates the existence of cold dark matter.
Naturally, the validity of the $\Lambda$CDM model, along with its fundamental principles, is continually challenged, and any necessary modifications are made with the aim of overcoming these difficulties with minimal adjustments.
However, we anticipate an increase in this debate in the near future, driven by the forthcoming results from the James Webb Space Telescope.

Since a small nonvanishing cosmological constant is required in the $\Lambda$CDM model, predicting or deriving the observed value of the cosmological constant remains a challenging problem.
Although numerous proposals attempt to explain the smallness of the cosmological constant, aiming to solve this fine-tuning problem, its fundamental origin has not been elucidated \cite{Weinberg:1988cp,Padmanabhan:2002ji,Bousso:2007gp,Padilla:2015aaa}.
At the classical level, there is no issue with having such a tiny value for the cosmological constant.
The problem arises when quantum corrections, contributing to the vacuum energy, are included for the cosmological constant.
These corrections could significantly increase its value, rendering the cosmological constant unstable to large quantum corrections.
This poses an important challenge because many cosmological scenarios strongly depend on the cosmological constant.
Without a definite origin for this constant, these scenarios also suffer from a lack of deeper understanding.

The G\"{o}del Universe is one of the many scenarios that depend on the presence of the cosmological constant in general relativity (GR) \cite{Godel}.
It is an exact cosmological solution of Einstein's field equations with rotating matter, known for the possibility of the presence of Closed Timelike Curves (CTCs) that can lead to the violation of causality.
This violation is not local, as spacetimes in general relativity are locally equivalent to the causal structure of special relativity; hence, causality violation is globally allowed.
To examine the causality violation in more detail, a generalization of this metric, called the G\"{o}del-type metric, was developed \cite{tipoGodel}.
In this context, depending on the values assumed by the theory's parameters, three different classes of the G\"{o}del-type metric can be analyzed: linear, trigonometric and hyperbolic.
In addition, each one of these classes leads to a different causal structure for the spacetime; in fact, causal and non-causal regions can be realized.
These metrics have been examined for their consistency beyond general relativity, as well as the possibility of causal solutions in some modified gravity models  \cite{Clifton:2005at,Agudelo:2016pic,Porfirio:2016nzr,Porfirio:2016ssx,Agudelo:2016pic,Gama:2017eip,Nascimento:2021bzb,Altschul:2021rog,Luminet:2021qae,Nascimento:2023zok}.

Based on the above remarks, it is interesting to discuss the causal structure of the G\"{o}del-type universe within a theory where the cosmological constant plays a prominent role, offering a different perspective on its origin.
This is precisely the case of unimodular gravity (UG).
Initially, Einstein considered the unimodular condition $\sqrt{-g}=1$ \cite{Einstein1} as a convenient way to partially (gauge) fix a coordinate system in GR, simplifying calculations in certain situations.
Later on, this approach was realized as an alternative theory of gravity \cite{Einstein2}.
This definition of unimodular gravity is based on the invariance under a restricted group of diffeomorphisms, called transverse diffeomorphisms, which leave the determinant of the metric invariant \cite{Bufalo,Padilha,Comparison,Jirousek:2023gzr,Alvarez:2023utn,Bengochea:2023dep}.
Since the unimodular framework has a different symmetry content than GR, the cosmological constant in this theory appears as a constant of integration in the field equations, rather than a coupling constant.
Therefore, as the value of the cosmological constant is unspecified and unrelated to any coupling constant, problems associated with the cosmological constant can be reconsidered from a new perspective \cite{Comparison,Jirousek:2023gzr,Alvarez:2023utn,Bengochea:2023dep}.

On one side, regarding the quantum equivalence between UG and GR, a considerable amount of work has been done to determine whether the cosmological constant fine-tuning problem (due to matter loops) is also present in unimodular gravity, but without reaching a conclusive answer \cite{Bufalo,Padilha,Saltas:2014cta,Comparison, Kaloper:2013zca,Kaloper:2015jra,Alvarez:2015pla,Alvarez:2015sba,fR,Percacci:2017fsy,deBrito:2021pmw,string, BD,Klinkhamer:2022mxo}.
Another aspect worth mentioning is that at the quantum level, the unimodular condition can manifest itself in different ways, as well as influencing the interpretation of $\Lambda$ \cite{Bufalo}.
In the path integral and the quantum effective
action of the ordinary unimodular theory, the unimodular condition is averaged over space \cite{Bufalo}.
In the Henneaux-Teitelboim formulation \cite{Henneaux:1989zc}, the unimodular condition is locally imposed in the path integral as well as in the quantum effective action \cite{Smolin:2009ti}.
In general, the unimodular path integral, in any of these formulations, differs from the path integral of GR and might lead to different predictions. For a detailed account, see the references \cite{Bufalo,Padilha,Comparison,Jirousek:2023gzr,Alvarez:2023utn}.

On the other hand, the construction of metric solutions in the unimodular gravity is subtle, mainly because is necessary to cast the metric in the form $\sqrt{-g}=1$.
This process, as is must be emphasized, should not be taken as corresponding to a
standard change of coordinates, but rather to a new solution, in different coordinates \cite{Alvarez:2023utn,Bengochea:2023dep}. 
This observation, that one should write the metric in the appropriated coordinates, unveils several new aspects of metric solutions in the unimodular approach, for more details we mention to the refs. \cite{Alvarez:2023utn,Bengochea:2023dep} and references therein.
Moreover, although the field equations of unimodular gravity are classically equivalent to those of GR with an unspecified cosmological constant, \footnote{It is important to remark that this equivalence is no longer true beyond the background level when perturbations are considered \cite{gao2014cosmological,Alva}.}
 because the cosmological constant has different roles in these theories, one might expect that they can have a different impact on G\"{o}del-type solutions.
Furthermore, since the causal structure of the G\"{o}del-type universe is sensitive to the presence of the cosmological constant, one should expect that the unimodular theory can shed new light on the issue of causality violation and perhaps modify acausal regions of the solution compared to those of GR.

Hence, the main purpose of this paper is to present a detailed analysis of causality and its violation in the G\"{o}del-type universe within the unimodular theory. 
The interest in this proposal is twofold: (i) in GR, the G\"{o}del metric requires the cosmological constant to assume a special value; (ii) in UG, the cosmological constant has a new interpretation.
Therefore, it is of physical importance to verify whether this solution is allowed in UG and whether the condition imposed upon the cosmological constant is the same as in GR.
Actually, we obtain that in general the G\"odel solution in the unimodular framework is not fully equivalent to GR unless a special choice for the unimodular metric is considered.
A non-equivalence of some metric solutions between UG and GR has been observed in other cases \cite{Alvarez:2023utn}.

The present paper is organized as follows. In section \ref{sec2}, a brief introduction to unimodular gravity is presented.
In section \ref{sec3}, the original G\"{o}del universe is considered and the GR field equations are solved for rotating dust as matter content.
When the original G\"{o}del metric is cast in an unimodular fashion, the resulting field equations do not possess a consistent solution for the (vorticity) parameter $a =a\left(\rho,\Lambda\right)$.
In order to have a better understanding of the absence of this solution in the unimodular framework, the G\"{o}del-type universe is considered in section \ref{sec4}.
It is shown the existence of solutions for the unimodular theory in the hyperbolic class of the G\"{o}del-type metric and that depending on the choice of matter content, causal and non-causal regions are found.
Additionally, we establish a comparison between the solutions of UG and GR, highlighting aspects related to the cosmological constant.
In section \ref{sec5}, remarks and conclusions are presented.

\section{Unimodular Gravity}
\label{sec2}

In this section, a brief review of the unimodular Gravity (UG) is presented.
UG is a modification of Einstein's theory of General Relativity (GR), widely considered the most accurate description of gravity and provides the foundation of the cosmological standard model ($\Lambda$CDM).
The unimodular approach is usually based on the invariance under a restricted group of diffeomorphisms that leave the determinant of the metric invariant.
This aspect allows setting the determinant of the metric tensor equal to a fixed constant $\xi$, ensuring that the overall scale of the spacetime is arbitrary \cite{Einstein1, Einstein2}.
The action describing this gravitational theory, where the unimodular condition $\sqrt{-g}=\xi $ is introduced as a constraint with a Lagrange multiplier $\chi$, is given by
\begin{equation}\label{action_uni}
    \mathcal{S}=\frac{1}{16\pi G}\int{d^4x\Big(\sqrt{-g}R-\chi(\sqrt{-g}-\xi)}\Big)+\int{d^4x\sqrt{-g}\mathcal{L}_m},
\end{equation}
where  $g$ is the determinant of the metric tensor $g_{\mu\nu}$, $R$ is the Ricci scalar curvature, and $\mathcal{L}_m$ is the Lagrangian density of matter.
Moreover, due to the presence of the fixed volume element $\xi \,d^4x$, the invariance under the (full) diffeomorphism ($x'^{\mu}(x)=x^{\mu}+\lambda^{\mu}$) exhibited in GR is lost, remaining only the invariance under transverse diffeomorphisms or volume-preserving diffeomorphisms ($\nabla_\mu \lambda^\mu =0$) \cite{Bufalo}. 

An unrestricted variation of the action \eqref{action_uni} with respect to the metric yields the following field equations
\begin{equation}\label{efe_uni}
    R_{\mu\nu}-\frac{1}{2}g_{\mu\nu}R+\frac{\chi}{2}g_{\mu\nu}=8\pi G T_{\mu\nu},
\end{equation}
in which $T_{\mu\nu}$ is the energy-momentum tensor of matter defined as
\begin{equation}
T_{\mu\nu}=-\frac{2}{\sqrt{-g}}\frac{\delta (\sqrt{-g}\mathcal{L}_m)}{\delta g^{\mu\nu}}.
\end{equation}
The variation of the action \eqref{action_uni} with respect to the auxiliary field $\chi$ gives the unimodular condition  $\sqrt{-g}=\xi$.
The trace of  Eq. \eqref{efe_uni} implies the following restrictive equation for the Lagrange multiplier
\begin{equation}\label{chi_uni}
    \chi=\frac{R}{2}+4\pi G \mathbb{T},
\end{equation}
where $\mathbb{T}=g^{\mu\nu}T_{\mu\nu}$ is the trace of the energy-momentum tensor.
Imposing the condition \eqref{chi_uni} upon Eq. \eqref{efe_uni}, it becomes the traceless form of Einstein's field equations:
\begin{equation}\label{efe_uni_final}
    R_{\mu\nu}-\frac{1}{4}g_{\mu\nu}R=8\pi G\left(T_{\mu\nu}-\frac{1}{4}g_{\mu\nu}\mathbb{T}  \right).
\end{equation}
Without loss of generality, the matter can be assumed to be diffeomorphism-invariant, satisfying $\nabla^\nu T_{\mu\nu}=0$, \footnote{There are also studies considering cosmological implications of nonconservative unimodular models \cite{Fabris:2021atr}.} leading to the relation
\begin{equation}\label{bianchi_identity}
\nabla^{\nu}R=-8\pi G\nabla^{\nu}\mathbb{T},
\end{equation}
where the Bianchi identities were used. This equation has the solution
\begin{equation}\label{uni_cosm_const}
    R=-8\pi G \mathbb{T}-4\Lambda,
\end{equation}
with $\Lambda$ being an integration constant that can be associated with the cosmological constant.
Using the result \eqref{uni_cosm_const} into the field equations \eqref{efe_uni_final}, the Einstein equations with a cosmological constant is recovered, justifying the above identification of the integration constant $\Lambda$.
Actually, it is precisely Eq. \eqref{uni_cosm_const} that highlights the main conceptual difference between GR and UG.
In UG, which is restricted to use coordinate systems that satisfy the constraint $\sqrt{-g}=\xi $, the cosmological constant appears as a constant of integration rather than a coupling constant, as in GR.

\section{G\"{o}del universe: Unimodular gravity vs general relativity}
\label{sec3}

It is well known that the causal anomaly, due to the presence of closed timelike curves (CTCs), \footnote{  An observer traveling along closed timelike curves (world line in a Lorentzian manifold) could return to its starting point, i.e., the past, and thus violate causality. The conditions for the emergence of such curves are imposed by the metric, as discussed in G\"{o}del's original paper \cite{Godel}.} is regarded as the main feature of the G\"{o}del universe.
Thus, based on the importance of the cosmological constant in G\"{o}del's solution in GR, one can naturally ask how this anomaly behaves in the unimodular theory since $\Lambda$ has a different nature in UG.
Hence, the next sections are devoted to examine aspects involving causality in the G\"{o}del universe within the unimodular gravity.

\subsection{General relativity}

The G\"{o}del metric represents a rotating, homogeneous and isotropic (but non-expanding) solution to Einstein equations of GR \footnote{ On the occasion of Einstein's 70th birthday in 1949,  K. G\"odel proposed an exact solution to Einstein's field equations, presenting local homogeneity of spacetime \cite{Godel}. }, which is expressed as  \cite{tipoGodel}
\begin{equation}
ds^{2}	=a^{2}\left(\left(dt+\mathcal{H}\left(x\right)dy\right)^{2}-dx^{2}-\mathcal{D}\left(x\right)dy^{2}-dz^{2}\right),
\end{equation}
where $\mathcal{H}\left(x\right)=e^{x}$ and $\mathcal{D}\left(x\right)=\frac{1}{2}e^{2x}$, or equivalently
\begin{equation}\label{GU_metric}
    ds^2=a^2(dt^2+2e^xdtdy+(e^{2x}/2)dy^2-dx^2-dz^2),
\end{equation}
in which $a$ is a positive number related with the angular velocity (vorticity) of the surrounding matter about the $y$-axis \cite{Godel}.

As usual, in order to solve the field equations in the G\"{o}del universe, some geometric quantities must be calculated, including the non-zero Ricci tensor components
\begin{equation}\label{GU_ricci}
    R_{00}=1,\quad  R_{22}=e^{2x}, \quad R_{02} =e^{x},
\end{equation}
and using $R=g^{\mu\nu}R_{\mu\nu}$, the Ricci scalar is simply given as
\begin{equation}\label{GU_ricci_scalar}
    R=1/a^2.
\end{equation}
Now, rotating dust is considered as the matter content, and its energy-momentum tensor is defined as
\begin{equation}\label{rotating_dust}
    T_{\mu\nu}=\rho ~ u_\mu u_\nu,
\end{equation}
where $\rho$ is the energy density and $u$ is a unit vector with components given by
\begin{equation}\label{GU_4velo_cov}
    u^{\mu}=\begin{pmatrix}
1/a, & 0, & 0, & 0 \end{pmatrix}, \quad u_{\mu}=\begin{pmatrix}
a, & 0, & ae^{x}, & 0 \end{pmatrix},
\end{equation}
which satisfy the condition $u_{\mu}u^{\mu}=1$.
Finally, using Eq. \eqref{GU_4velo_cov}, the covariant components of the energy-momentum tensor \eqref{rotating_dust} are given as
\begin{equation}\label{GU_energy_tensor}
   T_{\mu\nu}=\begin{pmatrix}
\rho a^2 & 0 & \rho a^2e^x & 0 \\
0 & 0 & 0 & 0 \\
\rho a^2e^x & 0 & \rho a^2e^{2x} & 0 \\
0 & 0 & 0 & 0 
\end{pmatrix}.
\end{equation}

With these results, the GR field equations with a cosmological constant
\begin{equation} \label{fe_GR}
 R_{\mu\nu}-\frac{1}{2}Rg_{\mu\nu}- \Lambda g_{\mu\nu}=8\pi GT_{\mu\nu},
\end{equation}
can be solved for the G\"{o}del universe \eqref{GU_metric} and a necessary condition upon the cosmological constant arises
\begin{equation} \label{GU_cosm_const}
   - \Lambda = \frac{1}{2a^2}=4\pi G\rho.
\end{equation}
This result identifies the so-called G\"odel metric as a solution of Einstein's equations with a (negative) cosmological constant $\Lambda$ supported by a dust of density $\rho$ \footnote{Although the anti de-Sitter (AdS) model also presents a negative cosmological constant, G\"odel universe is characterized as being sourced by a rotating dust while AdS is sourced by vacuum }. 
Furthermore, since the existence of CTCs is allowed in the G\"{o}del universe, causality is violated in this solution.

\subsection{Unimodular gravity}

Let us now consider the (original) G\"{o}del metric \eqref{GU_metric} with a rotating dust in the UG framework.
The first step in analyzing the unimodular theory is to rewrite the metric \eqref{GU_metric} in order to satisfy the unimodular condition $\sqrt{-g}=1$ (with a fixed density $\xi=1$, for simplicity).
Hence, the condition $\sqrt{-g}=1$ can be achieved by means of a change in the time coordinate $t\to t'=a^{-4}e^{-x}t$, so that the metric \eqref{GU_metric} is cast in a unimodular fashion as
\begin{equation} \label{eq_new_unid}
ds^{2}=a^{-6}e^{-2x}dt^{2}+2a^{-2}dtdy-a^{2}\sqrt{2}dx^{2}+a^{2}\frac{e^{2x}}{2}dy^{2}-a^{2}\sqrt{2}dz^{2}.
\end{equation}
In this case, the non-zero Ricci tensor components are
\begin{align}
R_{00}  =\frac{2\sqrt{2}e^{-2x}}{a^{8}},\quad R_{02}  =\frac{\sqrt{2}}{a^{4}},\quad 
R_{11}  =2,\quad R_{22}   =\sqrt{2}e^{2x},
\end{align}
while the Ricci scalar reads
\begin{align}
R=-\frac{\sqrt{2}}{a^{2}}.
\end{align}
Moreover, in terms of the metric \eqref{eq_new_unid} the energy-momentum tensor for the rotating dust \eqref{rotating_dust} is
\begin{align}
T_{\mu\nu}=\left(\begin{array}{cccc}
\frac{\rho e^{-2x}}{a^{6}} & 0 & \frac{\rho}{a^{2}} & 0\\
0 & 0 & 0 & 0\\
\frac{\rho}{a^{2}} & 0 & a^{2}\rho e^{2x} & 0\\
0 & 0 & 0 & 0
\end{array}\right)
\end{align}
and the trace of the energy-momentum tensor is simply $\mathbb{T}=\rho$.

Hence, from the above definitions, the field equations \eqref{efe_uni_final} are written as
\begin{align}
\frac{\sqrt{2}}{a^{2}}&=\frac{8}{3}\pi G\rho\\
\frac{\sqrt{2}}{a^{2}}&=\frac{24}{5}\pi G\rho\\
\frac{\sqrt{2}}{a^{2}}&=\frac{56}{9}\pi G\rho
\end{align}
and the condition \eqref{uni_cosm_const} for the cosmological constant $\Lambda$ leads to
\begin{equation}
\frac{\sqrt{2}}{a^{2}}=8\pi G\rho+4\Lambda.
\end{equation}
We thus observe that no solution of $a =a\left(\rho,\Lambda\right)$ can satisfy all the above relations.
In summary, the present analysis indicates that, contrary to what happens in GR, the (original) G\"{o}del universe is not a solution in the unimodular framework, as the field equations cannot be solved (this conclusion remains valid if fluid pressure is added to the source (a perfect fluid)).

To better understand the `limitation' that the unimodular condition imposes upon the (original) G\"{o}del metric and explore the physical invariance under coordinate transformations, we will consider a generalized version of this metric in the next section.
With this new class, named G\"{o}del-type metrics \cite{tipoGodel}, we have a higher symmetry content (as well as parameters), which allows a suitable discussion of the unimodular condition.
Combined with different types of matter content, these aspects allow us to investigate how the unimodular condition affects the G\"{o}del solution and also how the equivalence between GR and UG is preserved within the G\"{o}del universe.

\section{G\"{o}del-type Metrics: Unimodular gravity vs general relativity}
\label{sec4}

The (original) G\"{o}del metric, given in Eq. \eqref{GU_metric}, represents a homogeneous space-time in rotation.
This metric stimulated the investigation of more complex solutions describing rotating cosmological spacetimes.
Over the years, several metrics, namely G\"{o}del-type metrics, \footnote{For a historical review of studies that lead to the G\"{o}del-type metrics see the references \cite{Ozsvath, Banerjee, Som, Raychaudhuri, Bampi}.
} have been examined.
Obviously, the set of homogeneous G\"{o}del-type space-times is not exhausted, and solutions to Einstein equations can be obtained by assuming different matter contents.

Here the G\"{o}del-type metric proposed in ref.~\cite{tipoGodel} is considered, where the viability of G\"{o}del-type universes within the unimodular theory is examined.
In this case, causal and non-causal regions arise for different types of matter contents.
Therefore, it is possible to draw a relationship between the violation of causality and the considered content of matter.
Additionally, the equivalence of GR and UG is discussed in terms of their solutions for G\"{o}del-type universes.

In this context, the G\"{o}del-type metric is considered in cylindrical coordinates $(r,\phi, z)$ given by \cite{tipoGodel}
\begin{equation}\label{metric_godel_type1}
ds^{2}=\left[dt+H(r)d\phi\right]^{2}-D^{2}(r)d\phi^{2}-dr^{2}-dz^{2},
\end{equation}
with $H(r)$ and $D(r)$ being arbitrary metric functions that satisfy the conditions
\begin{align}
 \frac{H'(r)}{D(r)}&=2\omega, \label{cond1}\\
 \frac{D''(r)}{D(r)}&=m^2,\label{cond2}
\end{align}
where the prime denotes the derivative with respect to $r$.
These conditions arise when the Killing equations, associated with the Riemannian manifold described by the homogeneous metric \eqref{metric_godel_type1}, are solved.
In ref.~\cite{tipoGodel} the general problem of finding the Killing vectors of all Riemannian G\"{o}del-type manifolds was extensively examined. 

The parameters $m^2$ and $\omega$ are constants that completely characterize the properties of the (homogeneous) G\"{o}del-type metrics.
These parameters take the values: $-\infty\leq m^2\leq\infty$ and $\omega\neq 0$, as $\omega$ is identified as the vorticity.
Interestingly, depending on the sign of $m^2$ and $\omega\neq 0$, there are three different classes of G\"{o}del-type metrics \cite{tipoGodel}: (i) hyperbolic class ($m^2>0$), (ii) trigonometric class ($m^2=-\mu^2<0$), and (iii) linear class ($m^2=0$).
Furthermore, since the G\"{o}del metric \eqref{GU_metric} is recovered for $m^2=2\omega^2$, one can conclude that it belongs to the hyperbolic class.

In order to examine the CTCs, the G\"{o}del-type metric \eqref{metric_godel_type1} can be expressed as
\bea
ds^{2}=dt^2+2H(r)dt d\phi - W(r)d\phi^{2}-dr^{2}-dz^{2},
\eea
with $W(r)=D^2(r)-H^2(r)$.
The most interesting feature of this metric is that it presents CTCs defined by circles $\mathcal{C} = \left\{ (t,r,\phi,z);~ (t,r,z=\mathrm{const}), ~\phi\in [0,2\pi] \right\}$ and depend strongly on the behavior of the function $W(r)$.
More specifically, CTCs arise only when  $W(r)<0$ in a region limited by the range ($r_1 < r < r_2$).

For the linear class $m = 0$, there exists a noncausal region $r > r_c$ possessing closed time-like curves, where
\begin{equation}
r_c = 1/\omega
\end{equation}
is the critical radius.	
Now, for the hyperbolic class, it can be separated into two subclasses based on the values of the parameter $m^2$:
(i) if the values of $m^2$ lie within $0<m^2<4\omega^2$,  then the spacetime possesses one non-causal region for $r > r_c$, with the critical radius $r_c$ determined by
\bea \label{crit_rad}
\sinh^2\left(\frac{mr_c}{2}\right)=\left(\frac{4\omega^2}{m^2}-1\right)^{-1}.
\eea
(ii) On the other hand, if $m^2 > 4\omega^2$, then there is no occurrence of CTCs, and the universe is completely causal.
At last, in the special case $m^2 = 4\omega^2$, the critical radius $r_c \to \infty$ \cite{tipoGodel} and no violation of causality is found. 

In order to analyse the G\"{o}del-type universes, for simplicity, we shall consider the vierbein formalism.
Let us define a local set of vierbein basis for the spacetime $\theta^{A}=e^A_{\phantom{A}\mu} dx^\mu$, in which $e^A_{\phantom{A}\mu}$ are called vierbein fields.
Thus, the line element \eqref{metric_godel_type1} assumes the form
\begin{equation}\label{metric_godel_type02}
ds^{2}=\eta_{AB}\theta^{A}\theta^{B},
\end{equation}
where
\begin{align}\label{vierbein1} 
 \theta^{(0)}=dt+H(r)d\phi;&\quad \theta^{(1)}=dr;\\ 
    \theta^{(2)}=D(r)d\phi; &\quad \theta^{(3)}=dz,\label{vierbein2}
\end{align}
and $\eta_{AB}=\text{diag}(1,-1,-1,-1)$ is the Minkowski metric.
Note that capital Latin letters are used to refer to the local Lorentz (co)frame, while small Greek label spacetime coordinates.
Considering the hyperbolic class, the functions $H(r)$ and $D(r)$ take the form
\begin{align}
 H(r)&=\frac{4\omega}{m^2}\sinh{^2\bigg(\frac{mr}{2}\bigg)},\\
 D(r)&=\frac{1}{m}\sinh{(mr)}.
\end{align}
This class is of particular interest as it exhibits a well-defined separation of causal and non-causal regions in terms of the critical radius $r_c$ \cite{tipoGodel}. On the other hand, the trigonometric class presents an infinite sequence of alternating causal and non-causal regions.

In terms of the vierbein basis, the field equations can be written without difficulty.
It is essential to use the vierbein $e^A_{\phantom{A}\mu}$ or its inverse $E^{\phantom{A}\mu}_A$ to convert the spacetime indices into Lorentz indices of tensor objects. For instance, the Ricci tensor is expressed as  $R_{(A)(B)} = E^{\phantom{A}\mu}_A E^{\phantom{B}\nu}_B R_{\mu\nu}$. \footnote{To avoid confusion, the notation $T_{AB}=T_{(A)(B)}$ is used explicitly to refer to components of tensors defined with local Lorentz indices.}

With these elements we can start our review of G\"{o}del-type solution \eqref{metric_godel_type1} in GR with the simple case of the energy-momentum tensor for the \textbf{rotating-dust} which, in the local frame, is given as
\bea
T_{AB}=\rho \,u_A u_B, \label{dust}
\eea
in which $u^A=e^A_0=\delta^A_0$ is the four velocity of the fluid defined in the comoving frame. 

In this case, the field equations of GR \label{fe_GR} read
\begin{align} \label{GR_dust}
3\omega^2-m^2&=8\pi G\rho+\Lambda,\\
\omega^2&=-\Lambda,\\
m^2-\omega^2&=-\Lambda.
\end{align}
This set of equations has as solutions
\begin{equation}\label{cosm_const_gtm_rd}
m^{2}=2\omega^{2}=-2\Lambda=8\pi G\rho,
\end{equation}
This is precisely the same result \eqref{GU_cosm_const} found for the case of G\"{o}del original solution.
Therefore, as we have noted above, the relation $2\omega^2=m^2$ leads to the G\"{o}del solution, corresponding to a non-causal universe (i.e. a finite critical radius, beyond which the causality is violated, can be calculated).

We now consider a \textbf{perfect fluid} which energy-momentum tensor is described in a local frame as
\bea \label{eq_per_fluid}
T_{AB}=(\rho+p)u_A u_B-p~\eta_{AB}.
\eea
Then the GR field equations that describe the G\"{o}del-type universe are
\begin{align}
3\omega^2-m^2&=8\pi G\rho+\Lambda,\\
\omega^2&=8\pi G p-\Lambda,\\
m^2-\omega^2&=8\pi G p-\Lambda.
\end{align}
These equations have as solution the relations
\begin{equation}\label{final_gtm_pf_gr}
    2\omega^2=m^2=8\pi G(\rho+p)
\end{equation}
and 
\begin{equation}\label{cosm_const_pf_gr_final1}
   -2\Lambda= m^2-16\pi Gp.  
\end{equation}
Once again the observe the relation $2\omega^2=m^2$.
Since $m^2 >0$ from the above solutions they belong to the hyperbolic class.
This observation implies that in GR the non-causal solution is not avoided by the presence of a perfect fluid, again a finite critical radius can be calculated.

We consider a third type of source defined as a combination between a \textbf{perfect fluid and a scalar field}.
This proposal follows from the idea of examining whether different types of matter source affect the equivalence between the unimodular approach and GR.

In this case, the energy-momentum tensor is defined as
\begin{equation} \label{82}
T_{AB}=T^{(M)}_{AB}+T^{(S)}_{AB},
\end{equation}
with $T^{(M)}_{\mu\nu}$ being the part associated with the perfect fluid, given in Eq.  \eqref{eq_per_fluid}, and $T^{(S)}_{\mu\nu}$ is the part associated with the scalar field which reads
\begin{equation}\label{GU_pfs_energyscalar2}             
T^{(S)}_{AB}=\phi_{;A}\phi_{;B}-\frac{1}{2}\eta_{AB}\eta^{CD}\phi_{;C}\phi_{;D},
\end{equation}
where semicolon denotes the covariant derivative.
For matter of simplicity, the scalar field is parametrized as $\phi=\epsilon z+\epsilon$, with $\epsilon={\rm const}$ \cite{tipoGodel} .
Then, the components of the energy-momentum tensor for the scalar field become simply
\begin{equation}\label{properties_pfs_gtm_gr_2}
    T^{(S)}_{(0)(0)}=-T^{(S)}_{(1)(1)}=-T^{(S)}_{(2)(2)}=T^{(S)}_{(3)(3)}=\frac{\epsilon^2}{2}.
\end{equation}
Under these considerations, the GR field equations imply the relations
\begin{align}
3\omega^2-m^2&=8\pi G\rho+4\pi G\,\epsilon^2+\Lambda,\label{85}\\
\omega^2&=8\pi Gp-4\pi G\,\epsilon^2-\Lambda,\label{86}\\
m^2-\omega^2&=8\pi Gp+4\pi G\,\epsilon^2-\Lambda,\label{87}
\end{align}
which lead to a causal G\"{o}del-type solution given by
\begin{align}
m^2&= 4\omega^2\label{91},\\
8\pi G\rho&=-8\pi Gp=-2\omega^2-\Lambda.\label{92}
\end{align}
Based upon our previous discussion, the condition Eq. \eqref{91} yields an infinite critical radius.
Hence, under this combined matter source, causality is not violated and the CTCs are not allowed. Furthermore, this result shows the viability of a causal G\"{o}del-type universe in GR as long as the source \eqref{82} is considered.

We proceed now to the study of the G\"{o}del-type universe in the unimodular framework, the field equations \eqref{efe_uni_final} and \eqref{uni_cosm_const} are examined by specifying the matter content.
In order to examine the equivalence between GR and UG we consider again three types of sources: rotating dust, perfect fluid, and perfect fluid plus a scalar field.

\subsection{Unimodular gravity}

The first step of the analysis is to express the G\"{o}del-type metric \eqref{metric_godel_type1} in an unimodular form.
In order to satisfy the unimodular constraint $\sqrt{-g}=1$ we cast the line element \eqref{metric_godel_type1} as
\begin{equation} \label{metric_godel_type2}
ds_{2}=dt^{2}+2H(r)dtd\phi- D(r)^{-1}dr^{2}-W(r)d\phi^{2}- D(r)^{-1}dz^{2}.
\end{equation}
which yields the following vierbein basis
\begin{align}\label{vierbein3} 
 \theta^{(0)}=dt+H(r)d\phi;&\quad \theta^{(1)}=D^{-\frac{1}{2}} dr;\\ 
    \theta^{(2)}=D(r)d\phi; &\quad \theta^{(3)}=D^{-\frac{1}{2}}dz,\label{vierbein4}
\end{align}
This form of the metric is valid for all values of the parameters $m$ and $\omega$.
For our choice of unimodular metric \eqref{metric_godel_type2} we shall see that the corresponding field equations depend on the functions $D(r)$ and $H(r)$.
Hence, we examine whether or not the physical solutions of the field equations are valid for every  $D(r)$ and $H(r)$ or only for a special class of solutions (i.e. for a limited parameters phase space region).
This consistency analysis of the field equations will be also addressed in terms of the equivalence between GR and UG.

It should be emphasized that although there are different functional forms to the unimodular G\"{o}del-type metric (which can be cast in a gauge equivalent way to the metric \eqref{metric_godel_type1}), our choice by the form \eqref{metric_godel_type2} is because we can immediately drawn a series of physical consequences about this type of universe when the unimodular condition is imposed.
Hence, we are motived to work with the metric \eqref{metric_godel_type2} because the physical content of the model can be readily discussed and also is pivotal to understand the equivalence of UG with GR in the G\"odel universe.

We now apply these considerations to different sources in order to examine the unimodular G\"{o}del-type metric \eqref{metric_godel_type2} in the unimodular gravity.

\subsubsection{Rotating-dust}
\label{uni_dust}

Since we will study the unimodular theory \eqref{metric_godel_type2} in the vierbein basis, we have defined previously the sources.
Hence, we recall that the case for the rotating-dust has an energy-momentum tensor $T_{AB}$ given by \eqref{dust}.
 
Let us now examine the unimodular field equations. The components of the UG field equations \eqref{efe_uni_final} for a G\"{o}del-type metric \eqref{metric_godel_type2} sourced by a rotating-dust \eqref{dust} are given by
\begin{align} \label{eq_rot_dust}
\left(10\omega^{2}-m^{2}\right) D(r) &=24 \pi G  \rho\\
\left(6\omega^{2}-m^{2}\right)D(r)&=8\pi G \rho \\
\left(3m^{2}-2\omega^{2}\right) D(r)& =8\pi G\rho. 
\end{align}
In turn, using the unimodular constraint eq. \eqref{uni_cosm_const}, it is found that the cosmological constant reads
\begin{equation}\label{sulution_unimod_gtm3}
\left(m^{2}-2\omega^{2}\right)D(r)=-8\pi G\rho-4\Lambda.
\end{equation}
From a direct analysis one can easily conclude that
\begin{equation}
2\omega^{2}=m^{2}
\end{equation}
and also that 
\begin{equation} \label{sol_UG}
m^{2}=2\omega^{2}=-2\Lambda D^{-1}(r)=4\pi G\rho D^{-1}(r)
\end{equation}

One then observe that the G\"odel-type metric \eqref{metric_godel_type2} is a solution of the unimodular gravity, and reproduces the same causal behavior of the original G\"odel universe $m^{2}=2\omega^{2}$ \cite{tipoGodel}.
This is a surprising result since the original  G\"odel metric was not a solution of the unimodular gravity.
Actually, only when a larger number of parameters is introduced one is able to obtain a mathematically consistent solution to the unimodular field equations.

There are some important remarks to be made about the set of equations \eqref{eq_rot_dust}-\eqref{sulution_unimod_gtm3}, as well as from the solutions \eqref{sol_UG}: 

\begin{itemize}

\item One can see from the UG solutions \eqref{sol_UG} that they differ of the GR results \eqref{cosm_const_gtm_rd}, precisely by the function $D(r)$, yielding a departure between the theories.

\item Moreover, in order to achieve the unimodular metric \eqref{metric_godel_type2} we have introduced new factors of the function $D(r)$ in $\theta^{(1)}$ and $\theta^{(3)}$ (there are other choices, but all of them lead to the same conclusion).
This fact led to the field equations to depend explicitly on the variable $r$ (in terms of the function $D(r)$).
In general, there are no problems with such dependence, but there is a situation of caution: if we assume energy density and cosmological constant to be constants, we arrive at conditions like $D(r) = {\rm const}$; and such conditions, restricting the radial coordinates, are clearly inconsistent since the radial coordinate must be a free parameter.

\end{itemize}

Fortunately, one can impose strong constraints upon the acceptable form(s) of the function $D(r)$ if one requires the equivalence between UG and GR solutions (i.e. that they share the same causal structure), and also that the aforementioned inconsistency of the field equations is removed.
Obviously, these requirements shall restrict the phase space of solutions, but at least we can define the UG consistently.
Hence, all these conditions can be satisfied for the simplest choice $D(r)=1$, which corresponds, in terms of the condition \eqref{cond2}, to the linear class of solutions $m=0$ of the G\"odel-like metrics.
Moreover, if the condition \eqref{cond1} is considered, this constraint also yields the following solution $H(r) \sim \omega \,r$.
In this case, there is a noncausal region $r>r_c$, which can be determined in terms of the critical radius $r_c = 1/\omega$.
Actually, these conclusions are valid for all set of solutions discussed in this section for other sources.

\subsubsection{Perfect fluid}

We now consider a perfect fluid as the gravitational source of the unimodular gravity in the G\"odel-like metric, in which its energy-momentum tensor is described by \eqref{eq_per_fluid} with trace given by $\mathbb{T}=\rho-3p$.

Thus, the field equations for the unimodular framework \eqref{efe_uni_final} and the condition \eqref{uni_cosm_const}, for the G\"{o}del-type universe \eqref{metric_godel_type2},  yield the relations
\begin{align}
\left(10\omega^{2}-m^{2}\right) D(r) &=24 \pi G \left(p+\rho \right)\\
\left(6\omega^{2}-m^{2}\right) D(r) &=8\pi G  \left(p+\rho \right) \\
\left(3m^{2}-2\omega^{2}\right) D(r) &=8\pi G  \left(p+\rho \right) \\ 
\left(m^{2}-2\omega^{2}\right) D(r) &=-4\Lambda -8\pi G \left(\rho  -3p\right)
\end{align}
From this set of relations we obtain
\begin{equation} \label{eq123}
m^{2}=2\omega^{2}=4\pi G \left(\rho +p\right) D^{-1}(r)
\end{equation}
and also that
\begin{equation}\label{eq123b}
-2\Lambda =  m^2D (r) - 16 \pi G\,p 
\end{equation}
We observe that this set of solutions \eqref{eq123} and \eqref{eq123b} for UG: i) is not equivalent to GR eqs.~\eqref{final_gtm_pf_gr} and \eqref{cosm_const_pf_gr_final1}, and ii) that the field equations have a potential inconsistency for the case when the sources, energy density, pressure and cosmological constant are constants.
All these problems are the same as we have previously discussed in section \ref{uni_dust}, where we have concluded that these issues are completely solved for the choice $D(r)=1$ (the solution for $H(r)$ also holds), i.e. when we go to the linear class $m=0$ of solutions.

\subsubsection{Perfect fluid plus scalar field}

We now approach the unimodular theory sourced by a combination between perfect fluid and a scalar field given by \eqref{82} and \eqref{GU_pfs_energyscalar2}.
This is part of our endeavour to determine how different types of matter source can induce a causal G\"{o}del solution, as well as affects the equivalence between the unimodular approach and GR.
In this case the energy-momentum components are given by \eqref{82} and \eqref{properties_pfs_gtm_gr_2}, where its trace reads $\mathbb{T}=\rho-3p+\epsilon^2 $.

Under these considerations, the G\"{o}del-type metric \eqref{metric_godel_type2} is examined for this matter content in UG. 
Hence,  the UG field equations \eqref{efe_uni_final} sourced by the energy-momentum tensor \eqref{82}  are given as 
\begin{align}
\left(10\omega^{2}-m^{2}\right) D(r) &=8 \pi G \left[ 3 \left(p+\rho \right) +\epsilon^2 D(r)  \right]\label{93}\\
\left(6\omega^{2}-m^{2}\right) D(r) &=8\pi G \left[  \left(p+\rho \right) -\epsilon^2 D(r)  \right] \label{94} \\
\left(3m^{2}-2\omega^{2}\right) D(r) &=8\pi G \left[  \left(p+\rho \right) +3\epsilon^2 D(r)  \right]\label{95}   
\end{align}
and the cosmological constant can be determined using Eq. \eqref{uni_cosm_const}, which yields
\begin{equation}\label{96}
\left(m^{2}-2\omega^{2}\right)  D(r) =-4 \Lambda -8\pi G \left[  \left(\rho -3p \right) +\epsilon^2  D(r)  \right]
\end{equation}
whose solutions read
\begin{equation} \label{sol1}
m^{2}=4\omega^{2}
\end{equation}
and
\begin{equation}\label{sol2}
8\pi G\rho=-8\pi Gp=-2\omega^2 D(r)-\Lambda.
\end{equation}

Once again we notice that the solutions \eqref{sol1} and \eqref{sol2} for UG are not equivalent to GR eqs.~\eqref{91} and \eqref{92}, and also that the field equations have a potential inconsistency for the case when the sources are constants (due to the presence of the factor $D(r)$), which are handled for the choice $D(r)=1$, i.e. when we restrict to the linear class $m=0$ of solutions.

We have seen from the results of this section that the unimodular theory, independently of the source content, can be physical and mathematical consistently defined and be equivalent to GR only in a restricted part of the phase space of solutions,   valid for a special class of the functions $D(r)$ and $H(r)$, that corresponds to the linear class $m=0$ of solutions.
For completeness, we present a summary on the causal structure of GR and UG for different type of matter content in table \ref{tableI}.

\begin{table}

\begin{tabular}{c|c|c|c|}
\cline{2-4}
 & rotating dust & perfect fluid & perfect fluid and a scalar field\tabularnewline
\hline 
\multicolumn{1}{ |c| }{General Relativity}   & non-causal & non-causal & causal\tabularnewline
\hline 
\multicolumn{1}{ |c| }{Unimodular Gravity} & non-causal & non-causal & causal\tabularnewline
\hline 
\end{tabular}

\caption{A comparison summary about the causal structure of GR and UG.}
\label{tableI}
\end{table}

\section{Conclusion}
\label{sec5}

It was examined in this paper physical aspects of the G\"{o}del-type universes in the unimodular gravity.
A key point for the existence of G\"{o}del solutions is the presence of a cosmological constant in GR, which usually is added by hand as a coupling constant.
This very fact leads to a significant interest in how the G\"{o}del solutions behave in the unimodular theory, since this setting naturally presents a cosmological constant in its structure and, therefore, UG could be a suitable framework to examine G\"{o}del-type universes.
Moreover, since the most appealing feature of G\"{o}del universes is the possibility of causality violation through the existence of CTCs, it is very natural to ask whether UG and GR lead to physically equivalent solutions.

In addition, it is important to emphasize that if gravitation interaction is governed by the unimodular theory rather than general relativity, various issues should be reexamined in this framework.
An important issue includes the breakdown of causality described by G\"{o}del-type universes, which we explore in the unimodular theory, and a comparison between the two theories is carried out where the conditions for obtaining causal and non-causal regions is detailed. 

Different types of gravitational sources, rotating dust, perfect fluid and perfect fluid plus a scalar field, are considered in our analysis in order to examine whether and how they can influence the causal structure of the G\"{o}del-type solution.
Moreover, since the field equations in the unimodular setting depend explicitly on the function $D(r)$, the requirements that the physical solutions were equivalent to GR and that the field equations were mathematically consistent imposed strong constraints upon the function $D(r)$.
Hence, we have shown that in order to satisfy all these conditions the choice $D(r)=1$ is the more appropriated, showing that the set of physical solutions is restricted only to a special class of functions $D(r)$, i.e. to the linear class $m=0$ of solutions.
Moreover, if the condition \eqref{cond1} is considered, this constraint also yields the following solution $H(r) \sim \omega \,r$.
In this case, all the solutions present a non-causal region $r>r_c$, where the critical radius is $r_c = 1/\omega$.

At last, since the cosmological constant has a natural emergence in the unimodular theory, one could naively think that it would have a closer relation with the G\"{o}del and G\"{o}del-type universes, our results establishing the equivalence shows that the G\"{o}del solutions is indifferent either GR or UG is considered as a dynamical theory.
Moreover, a tempting continuation of the present study is to consider a general framework of the $F(R)$ theory and by the process of reconstruction determine functional forms of the $F(R)$ that possess G\"odel and G\"odel-type universes as solutions belonging to the hyperbbolic or trigonometric classes (i.e. $D(r) \neq {\rm const}$).

\section*{Acknowledgments}

This work by A. F. S. is partially supported by National Council for Scientific and Technological Development - CNPq project No. 312406/2023-1. R. R. thanks CAPES for financial support.
R.B. acknowledges partial support from Conselho
Nacional de Desenvolvimento Cient\'ifico e Tecnol\'ogico (CNPq Project No. 306769/2022-0). 

\section*{Data Availability Statement}

No Data associated in the manuscript.

\section*{Conflict of interest}

Authors declare that there is no conflict of interests.


\global\long\def\link#1#2{\href{http://eudml.org/#1}{#2}}
 \global\long\def\doi#1#2{\href{http://dx.doi.org/#1}{#2}}
 \global\long\def\arXiv#1#2{\href{http://arxiv.org/abs/#1}{arXiv:#1 [#2]}}
 \global\long\def\arXivOld#1{\href{http://arxiv.org/abs/#1}{arXiv:#1}}



\begin{thebibliography}{99}



\bibitem{Weinberg:1988cp}
S.~Weinberg,
``The Cosmological Constant Problem,''
\doi{10.1103/RevModPhys.61.1}{Rev. Mod. Phys. \textbf{61} (1989), 1-23}

\bibitem{Padmanabhan:2002ji}
T.~Padmanabhan,
``Cosmological constant: The Weight of the vacuum,''
\doi{10.1016/S0370-1573(03)00120-0}{Phys. Rept. \textbf{380} (2003), 235-320}


\bibitem{Bousso:2007gp}
R.~Bousso,
``TASI Lectures on the Cosmological Constant,''
\doi{10.1007/s10714-007-0557-5}{Gen. Rel. Grav. \textbf{40} (2008), 607-637}

\bibitem{Padilla:2015aaa}
A.~Padilla,
``Lectures on the Cosmological Constant Problem,''
\arXiv{1502.05296}{hep-th}.




\bibitem{Godel} K. G\"{o}del, ``An Example of a New Type of Cosmological Solutions of Einstein's Field Equations of Gravitation',' 
\doi{10.1103/RevModPhys.21.447} {Rev. Mod. Phys. {\bf 21}, 447 (1949).}

\bibitem{tipoGodel} M. J. Rebou\c{c}as and J. Tiomno, ``Homogeneity of Riemannian space-times of G\"{o}del type','
\doi{10.1103/PhysRevD.28.1251} {Phys. Rev. D {\bf 28}, 1251 (1983).}



\bibitem{Clifton:2005at}
T.~Clifton and J.~D.~Barrow,
``The Existence of godel, Einstein and de Sitter universes,''
\doi{10.1103/PhysRevD.72.123003}{Phys. Rev. D \textbf{72} (2005), 123003}

\bibitem{Agudelo:2016pic}
J.~A.~Agudelo, J.~R.~Nascimento, A.~Y.~Petrov, P.~J.~Porf\'\i{}rio and A.~F.~Santos,
``G\"odel and G\"odel-type universes in Brans\textendash{}Dicke theory,''
\doi{10.1016/j.physletb.2016.09.011}{Phys. Lett. B \textbf{762} (2016), 96-101}

\bibitem{Porfirio:2016nzr}
P.~J.~Porfirio, J.~B.~Fonseca-Neto, J.~R.~Nascimento, A.~Y.~Petrov, J.~Ricardo and A.~F.~Santos,
``Chern-Simons modified gravity and closed timelike curves,''
\doi{10.1103/PhysRevD.94.044044}{Phys. Rev. D \textbf{94} (2016) no.4, 044044}

\bibitem{Porfirio:2016ssx}
P.~J.~Porfirio, J.~B.~Fonseca-Neto, J.~R.~Nascimento and A.~Y.~Petrov,
``Causality aspects of the dynamical Chern-Simons modified gravity,''
\doi{10.1103/PhysRevD.94.104057}{Phys. Rev. D \textbf{94} (2016) no.10, 104057}
 
\bibitem{Gama:2017eip}
F.~S.~Gama, J.~R.~Nascimento, A.~Y.~Petrov, P.~J.~Porfirio and A.~F.~Santos,
``G\"odel-type solutions within the f(R,Q) gravity,''
\doi{10.1103/PhysRevD.96.064020}{Phys. Rev. D \textbf{96} (2017) no.6, 064020}

\bibitem{Nascimento:2021bzb}
J.~R.~Nascimento, A.~Y.~Petrov and P.~J.~Porf\'\i{}rio,
``Causal G\"odel-type metrics in non-local gravity theories,''
\doi{10.1140/epjc/s10052-021-09640-5}{Eur. Phys. J. C \textbf{81} (2021) no.9, 815}


\bibitem{Altschul:2021rog}
B.~Altschul, J.~R.~Nascimento, A.~Y.~Petrov and P.~J.~Porf\'\i{}rio,
``First-order perturbations of G\"odel-type metrics in non-dynamical Chern\textendash{}Simons modified gravity,''
\doi{10.1088/1361-6382/ac3e50}{Class. Quant. Grav. \textbf{39} (2022) no.2, 025002} 


\bibitem{Luminet:2021qae}
J.~P.~Luminet,
``Closed Timelike Curves, Singularities and Causality: A Survey from G\"odel to Chronological Protection,''
\doi{10.3390/universe7010012}{Universe \textbf{7} (2021) no.1, 12}.



\bibitem{Nascimento:2023zok}
J.~R.~Nascimento, A.~Y.~Petrov, P.~J.~Porfirio and R.~N.~da Silva,
``G\"odel-type solutions within the $f(R,Q,P)$ gravity,''
\arXiv{2302.09935}{gr-qc}.



\bibitem{Einstein1} A. Einstein, ``The Foundation of the General Theory of Relativity'', 
\doi{10.1002/andp.19163540702} {Annalen Phys. {\bf 49}, 769 (1916)}. Translated and included in The Principle of Relativity, by H.A. Lorentz et al. (Dover Press, New York, 1923).

\bibitem{Einstein2} A. Einstein, ``Do gravitational fields play an essential part in the structure of the elementary particles of matter?'', Sitzungsber. Preuss. Akad. Wiss. {Berlin (Math. Phys.) {\bf 1919}, 433 (1919).} Translated and included in The Principle of Relativity, by H.A. Lorentz et al. (Dover Press, New York, 1923).	


 

\bibitem{Bufalo} R. Bufalo, M. Oksanen and A. Tureanu, ``How unimodular gravity theories differ from general relativity at quantum level'',
\doi{10.1140/epjc/s10052-015-3683-3} {Eur. Phys. J. C {\bf 75}, 477 (2015).}




 \bibitem{Padilha} A. Padilla and I. D. Saltas, ``A note on classical and quantum unimodular gravity,''
\doi{10.1140/epjc/s10052-015-3767-0} {Eur. Phys. J. C {\bf 75}, 561 (2015).} 


\bibitem{Comparison} R. Carballo-Rubio, L. J. Garay and G. Garc\'ia-Moreno, ``Unimodular Gravity vs General Relativity: A status report'',
\doi{10.1088/1361-6382/aca386} {Class. Quantum Grav. {\bf 39}, 243001 (2022).}

 
\bibitem{Jirousek:2023gzr}
P.~Jirou\v{s}ek,
``Unimodular Approaches to the Cosmological Constant Problem,''
\doi{10.3390/universe9030131}{Universe \textbf{9} (2023) 131}.


\bibitem{Alvarez:2023utn}
E.~Alvarez and E.~Velasco-Aja,
``A Primer on Unimodular Gravity,''
 \arXiv{2301.07641}{gr-qc}. 
 
 \bibitem{Bengochea:2023dep}
G.~R.~Bengochea, G.~Leon, A.~Perez and D.~Sudarsky,
``A clarification on prevailing misconceptions in unimodular gravity,''
\doi{10.1088/1475-7516/2023/11/011}{JCAP \textbf{11} (2023), 011}
\arXiv{2308.07360}{gr-qc}. 
 
 
\bibitem{Saltas:2014cta}
I.~D.~Saltas,
``UV structure of quantum unimodular gravity,''
\doi{10.1103/PhysRevD.90.124052}{Phys. Rev. D \textbf{90} (2014) no.12, 124052}
 

\bibitem{Kaloper:2013zca}
N.~Kaloper and A.~Padilla,
``Sequestering the Standard Model Vacuum Energy,''
\doi{10.1103/PhysRevLett.112.091304}{Phys. Rev. Lett. \textbf{112} (2014) no.9, 091304}

 \bibitem{Kaloper:2015jra}
N.~Kaloper, A.~Padilla, D.~Stefanyszyn and G.~Zahariade,
``Manifestly Local Theory of Vacuum Energy Sequestering,''
\doi{10.1103/PhysRevLett.116.051302}{Phys. Rev. Lett. \textbf{116} (2016) no.5, 051302}

\bibitem{Alvarez:2015pla}
E.~\'Alvarez, S.~Gonz\'alez-Mart\'in, M.~Herrero-Valea and C.~P.~Mart\'\i{}n,
``Unimodular Gravity Redux,''
\doi{10.1103/PhysRevD.92.061502}{Phys. Rev. D \textbf{92} (2015) no.6, 061502}

\bibitem{Alvarez:2015sba}
E.~\'Alvarez, S.~Gonz\'alez-Mart\'in, M.~Herrero-Valea and C.~P.~Mart\'\i{}n,
``Quantum Corrections to Unimodular Gravity,''
\doi{10.1007/JHEP08(2015)078}{JHEP \textbf{08} (2015), 078}


\bibitem{fR} S. Nojiri, S. D. Odintsov and V. K. Oikonomou, ``Unimodular F(R) Gravity'',
\doi{10.1088/1475-7516/2016/05/046} { JCAP {\bf 05}, 046 (2016).}

 \bibitem{Percacci:2017fsy}
R.~Percacci,
``Unimodular quantum gravity and the cosmological constant,''
\doi{10.1007/s10701-018-0189-5}{Found. Phys. \textbf{48} (2018) no.10, 1364-1379}

\bibitem{deBrito:2021pmw}
G.~P.~de Brito, O.~Melichev, R.~Percacci and A.~D.~Pereira,
``Can quantum fluctuations differentiate between standard and unimodular gravity?,''
\doi{10.1007/JHEP12(2021)090}{JHEP \textbf{12} (2021), 090}


 \bibitem{string} L. J. Garay and G. Garc\'ia-Moreno, ``Embedding Unimodular Gravity in String Theory'',
 \doi{110.1007/JHEP03(2023)027} {	JHEP {\bf 03}, 027 (2023).}

 
\bibitem{BD} A. M. R. Almeida, J. C. Fabris, M. H. Daouda, R. Kerner, H. Velten and W. S. Hip\'olito-Ricaldi, ``Brans-Dicke unimodular gravity'',
\doi{10.3390/universe8080429c} {Universe {\bf 8}, 429 (2022).}

 


\bibitem{Klinkhamer:2022mxo}
F.~R.~Klinkhamer,
``Extension of unimodular gravity and the cosmological constant,''
\doi{10.1103/PhysRevD.106.124015}{Phys. Rev. D \textbf{106} (2022) no.12, 124015}



\bibitem{Henneaux:1989zc}
M.~Henneaux and C.~Teitelboim,
``The Cosmological Constant and General Covariance,''
\doi{10.1016/0370-2693(89)91251-3}{Phys. Lett. B \textbf{222} (1989), 195-199}

\bibitem{Smolin:2009ti}
L.~Smolin,
``The Quantization of unimodular gravity and the cosmological constant problems,''
\doi{10.1103/PhysRevD.80.084003}{Phys. Rev. D \textbf{80} (2009), 084003}
 \arXiv{0904.4841}{hep-th}. 
 

\bibitem{gao2014cosmological} C. Gao, R. H. Brandenberger, Y. Cai and P. Chen, ``Cosmological Perturbations in Unimodular Gravity'',
\doi{10.1088/1475-7516/2014/09/021} {JCAP {\bf 09}, 021 (2014).}



\bibitem{Alva} M. H. Alvarenga, J. C. Fabris and H. Velten,
``Using cosmological perturbation theory to distinguish between General Relativity and Unimodular Gravity'',
\doi{10.3390/sym15071392}{Symmetry \textbf{15} (2023) no.7, 1392}
 \arXiv{2301.12464}{gr-qc}. 
 
 \bibitem{Fabris:2021atr}
J.~C.~Fabris, M.~H.~Alvarenga, M.~Hamani-Daouda and H.~Velten,
``Nonconservative unimodular gravity: a viable cosmological scenario?,''
\doi{10.1140/epjc/s10052-022-10470-2}{Eur. Phys. J. C \textbf{82} (2022) no.6, 522}
 \arXiv{2212.06644}{gr-qc}. 

 
 \bibitem{Ozsvath} I. Ozsvath, ``New Homogeneous Solutions of Einstein's Field Equations with Incoherent Matter Obtained by a Spinor Technique'',
\doi{10.1063/1.1704311} {J. Math. Phys. {\bf 6}, 590 (1965).}

\bibitem{Banerjee} A. Banerjee and S. Banerji, ``Stationary distributions of dust and electromagnetic fields in general relativity'',
\doi{10.1088/0305-4470/1/2/302} {J. Phys. A {\bf 1}, 188 (1968).}

\bibitem{Som} M. M. Som and A. K. Raychaudhuri, ``Cylindrically symmetric charged dust distributions in rigid rotation in general relativity'',
\doi{10.1098/rspa.1968.0073} {Proc. R. Soc. London A {\bf 304}, 81 (1968).}

\bibitem{Raychaudhuri} A. K. Raychaudhuri and S. N. G. Thakurta, ``Homogeneous space-times of the G\"{o}del type'', 
\doi{10.1103/PhysRevD.22.802} {Phys. Rev. D {\bf 22}, 802 (1980).}

\bibitem{Bampi} F. Bampi and C. Zordan, ``A note on G\"{o}del's metric'', 
\doi{10.1007/BF00759840} {Gen. Relativ. Gravit. {\bf 9}, 393 (1978).}





 
 
 
 
 



  

 
\end{thebibliography}
\end{document}